\documentstyle[prd,epsf,aps]{revtex}
\begin{document}

\preprint{Imperial/TP/94-95/53 and {\tt hep-ph/9507423}}


\draft

\title{Vortex Production in Non-Relativistic and Relativistic
Media\footnote{Available in RevTeX format
through anonymous ftp from
{\tt ftp://euclid.tp.ph.ic.ac.uk/papers/94-5\_53.tex},
or on WWW in RevTeX and postscript formats at
{\tt http://euclid.tp.ph.ic.ac.uk/Papers/index.html} }}

\author{T.S. Evans\footnote{email: {\tt T.Evans@ic.ac.uk},
WWW: {\tt http://euclid.tp.ph.ic.ac.uk/$\sim$time}},
A.J. Gill\footnote{email: {\tt A.Gill@ic.ac.uk}} and
R.J. Rivers\footnote{email: {\tt R.Rivers@ic.ac.uk}}}

\address{Blackett Laboratory, Imperial College, South Kensington,
London SW7 2BZ, U.K.}
\date{26th July 1995}

\maketitle

\begin{abstract}
We examine string (vortex) formation
at a quench for a weakly-coupled global U(1) theory  when the
excitation spectrum is non-relativistic.    It  is so similar to
vortex production in the corresponding relativistic plasma as to
reinforce arguments for the similarity of vortex production in the
early universe and in low-temperature many-body physics.
\end{abstract}

\pacs{PACS Numbers : 2400}


\section{Introduction}

For many years it has been argued by Kibble \cite{kibble1} and
others \cite{shellard}
that the large-scale structure of
the universe can be attributed to cosmic strings formed during a
symmetry-breaking  transition at the Grand Unification scale.
Unfortunately, given the unlikely event of observing a cosmic string
(vortex) directly, chains of inference are sufficiently long that it
is difficult, if not impossible, to make the case compelling.

However,
the formation of topological defects like vortices during symmetry-breaking
phase
transitions is not unique to the early universe but generic to many physical
systems.
In particular, recent experiments on the
production
of vortices in superfluid $^{4}He$ \cite{lancaster} and $^{3}He$
\cite{helsinki}
have excited considerable interest.  The suggestion that these, and
other experiments \cite{lancaster2}, may provide an insight on the early
universe have been made by the experimental groups concerned.  In
this they have been championed by some theoretical astroparticle
physicists, most notably Zurek
\cite{zurek1} whose most recent review article \cite{zurek2} was, indeed,
titled
{\it Cosmological Experiments in Superfluids and Superconductors}.
Nonetheless, despite the hard thinking that has already taken place,
there is no doubt that a fuller
understanding of the {\it nonequilibrium} dynamics of vortex production is
required if comparisons are to be more than superficial analogies.

What encourages us in the hope that vortex production in the early
universe and the laboratory has close parallels is that, in the
first instance, the most plausible production mechanism refers to
neither.
We recapitulate it now for the simplest theory permitting vortices,
that of a complex scalar field $\phi ({\bf x},t)$.  The complex order
parameter of the theory is $\langle\phi\rangle = \eta e^{i\alpha}$ and the
theory possesses
a global $O(2)$ (or $U(1)$) symmetry that we take to be spontaneously broken
at a phase transition.  The field $\phi$ could
be either a complex non-relativistic order field appropriate to
a superfluid or a relativistic field in the early universe.
The transition is continuous in both cases but, as we shall see, the details of
the
transition order are largely irrelevant to our conclusions, as long
as it is not strongly first-order
\footnote{In that case the mechanism
for the transition, bubble nucleation, would lead to very different
consequences from those outlined below.}.

Initially, we take the system to be in the symmetry-unbroken
(disordered) phase.  We have no
reason to choose any particular initial field configuration, beyond
the requirement that the field is distributed about $\phi = 0$
with zero mean $\eta = |\langle\phi\rangle | =0$.
The simplest assumption is that, beginning at some time $t = t_0$,
the $O(2)$ symmetry of the ground-state (vacuum) is broken by a quench, a rapid
change in the
environment inducing an explicit time-dependence in the field
parameters. Once the quench is completed the $\phi$-field potential
$V(\phi ) = -a|\phi |^{2} + b|\phi |^{4}$ is
taken to have the symmetry-broken form $a > 0,\,\, b > 0$
of the familiar `wine-bottle' bottom.
The ground-state manifold
(the circle $S^{1}$, labelled by the phase $\alpha$ of $\langle\phi\rangle$) is
infinitely connected and the theory possesses {\it global} strings or
vortices, labelled by a winding number $n\in Z$.
Specific straight-string solutions to the classical field equations
are well-known, both for the non-relativistic \cite{kleinert} and the
relativistic cases \cite{shellard}.  The only property of these solutions that
we
need is that, as tubes of false vacuum, their thickness is the
Compton wavelength of the massive (Higgs) excitations of the theory.

These strings cost considerable energy to produce.  That they should appear
at all follows from a general argument,  due
to Kibble \cite{kibble1}, which goes as follows. During the
transition, the complex scalar field begins to
fall from the false ground-state into the true ground-state,
choosing a point on the ground-state manifold at each
point in space, subject to the constraint that it is
continuous and single-valued.
For continuous
transitions, for which this collapse to the true ground-state
occurs by spinodal decomposition or phase separation,
the resulting field configuration is expected to be one of
domains within each of which the scalar field has relaxed to a constant
ground-state value (i.e. constant field magnitude and phase).

If this is so, then the
requirements of continuity and single valuedness will sometimes force the field
to
remain in the false ground-state between some of the domains. For example, the
phase of
the field may change by an integer multiple of $2 \pi$ on going round
a loop in space. This requires at least one
{\it zero} of the field within the loop, each of which has topological
stability
and characterises a vortex passing through the
loop.   The density of strings is then closely linked to the number
of effective domains and the evolution of this density is,
correspondingly, linked to the nature of the domain growth.
When the phase transition is complete and there is no longer sufficient
thermal energy available for the field to fluctuate into the false
ground-state, the topological defects are frozen into the
field.
Simple counting arguments suggest defect densities at this time  of
the order of one string
passing through each correlation area.
{}From then on, the defect density alters almost entirely by
interactions of defects amongst themselves, rather than by
fluctuations in the fields \cite{tanmay}, but such calculations go
beyond the scope of this paper.

In an
earlier paper \cite{alray}, henceforth known as $I$, (but see also
\cite{timray}), two of us (A.G and R.R) showed how
defects were produced in a very simple model of a {\it relativistic}
$O(2)$ scalar field theory in which the long wavelength fluctuations
that drove the transition were Gaussian. The fluctuations were taken
to be set in train by the implementation of a quench from a high
temperature, above the transition, to a low temperature.
Despite its oversimplifications, its main interest
is that it provides a
concrete example in which the scenario outlined above is, indeed,
true and quantifiable.

With low temperature many-body theory in mind, in
this paper we show, in the same $O(2)$ model of Gaussian fluctuations,
how a density quench of a {\it nonrelativistic} medium
also leads to vortices. A priori, we might
expect significant differences in the vortex production.  The relativistic
regime
considered in $I$ was characterised by initial temperatures $T\gg m$
for all particle masses $m$ (in units $k_{B}=c=1$).  On the other
hand, non-relativistic media are characterised by $T\ll m$, to
freeze out antiparticles, but with $T\simeq \mu_{{\rm nr}}$, the
non-relativistic chemical potential.  Surprisingly, in our simple
model we find no difference  once a proper identification of mass
parameters has been made.

As it stands our model,
being weakly coupled (in effect, with zero-coupling for short times), is far
too simple
to mimic superfluid helium.
Equally, unless we are considering extremely weak couplings of the
strength invoked in inflationary models, our approximation will only
have limited applicability to hot quantum fields.  However, in
principle, if not yet in practice, we know how to incorporate
stronger interactions within our approximation scheme.  The identity
of the weak-coupling models should be reflected in similarities in
the more realistic stronger coupling counterparts.  As a result,
our main conclusion about the similarity of vortex
production in the two energy extremes may have greater generality,
even if the detailed results given here are too limited.

The organisation of this paper is as follows.  In $I$ we considered
the production of both $O(3)$ monopoles and $O(2)$ vortices.  Since
monopoles are somewhat easier to manage we concentrated on them at
the expense of vortices, for which we quoted only such properties as
were needed.  In the next section we repair this omission by
showing the way {\it vortex} densities, and density correlations,
can be inferred from field configuration probabilities through field
correlation functions.  As in $I$,
this summary is largely a recapitulation of some early work of
Halperin \cite{halperin}, modified to our purposes.

One of the difficulties of making straightforward comparisons
between early universe physics and many-body physics is the
difference in formalism, a reflection of the freezing out of
antiparticles in the non-relativistic regime.  We evade this problem
by showing how to incorporate both regimes
in a common relativistic expression, in which the non-relativistic
limit is obtained by tuning the chemical potential.
This permits the relevant field correlation functions
to be calculated in the same
approximations as for the relativistic case.

In the final sections we use these field correlation functions
as input for the vortex density functions as shown in section 2.  We conclude
with a
discussion of some of the results,
and consider their implications for numerical simulations.
In particular, the Kibble mechanism as presented above says nothing
about the fraction of strings that is produced in small loops
against the fraction of `infinite' string (i.e. string that does not
self-intersect).
Infinite string is necessary for the creation of large-scale
structure in the early universe.  We shall suggest that the fraction
of string in small loops is substantial.

\section{Vortex Distributions From Field Distributions}

Consider an ensemble of systems evolving from a given disordered
state
or, more realistically, from one of a set of disordered
states whose relative probabilities are known,
to an ordered state as indicated above, producing $O(2)$ vortices.
If the phase change begins
at time $t_{0}$ then, for $t > t_{0}$, it is  possible in principle
to calculate the probability $p_{t}[\Phi]$ that the complex field
$\phi ({\bf x}, t)$ takes the value $\Phi ({\bf x})$ at time $t$.
Throughout, it will be convenient to decompose $\Phi$ into
real and imaginary parts as $\Phi  = \frac{1}{\sqrt{2}}(\Phi_{1} +
i\Phi_{2})$ (and $\phi$ accordingly).  This is because we wish to
track the field as it falls from the unstable ground-state hump at the centre
of the potential to the ground-state manifold in
Cartesian field space.  It is trivial to reconvert the real
fields into complex fields (and similarly for conjugate fields, as
we need them).

The calculation of $p_{t}[\Phi]$ will be performed later in our simple model.
For the
moment, consider it given.  The question is, how can we infer the
string densities and the density correlations from  $p_{t}[\Phi]$?
That we can calculate them at all is a consequence of the fact,
noted earlier, that the
string core is a line of field zeroes.
This is equally true for both relativistic and non-relativistic
$O(2)$ theories
\footnote{However, this is not necessarily the case for the more
sophisticated vortices of $^{3}He$, for example, for which different
methods would be required.}.
The zeroes of $\Phi_{a}$ ($a$=1,2) which define the vortex positions
form either closed loops or the `infinite' string mentioned in the
introduction.
Following Halperin \cite{halperin} we define the {\it
topological line density} ${\vec \rho}(\bf r)$ by
\begin{equation}
{\vec \rho}({\bf r}) = \sum_{n}\int ds \frac{d{\bf R}_{n}}{ds}
\delta^{3} [{\bf r} - {\bf R}_{n}(s)].
\end{equation}
In (2.1) $ds$ is the incremental length along the line of zeroes ${\bf
R}_{n}(s)$ ($n$=1,2,.. .) and $\frac{d{\bf R}_{n}}{ds}$ is a unit
vector pointing in the direction which corresponds to positive
winding number.  Only winding numbers $n = \pm 1$ are considered.
Higher winding numbers are understood as describing multiple zeroes.
If $dA_j$ is an incremental two-dimensional surface containing the
point ${\bf r}$, whose normal is in the $j$th direction, then
$\rho_{j}({\bf r})$ is the net density of strings (i.e the density of
strings {\it minus} the density of antistrings on $dA_j$).

Ensemble averaging $\langle F[\Phi ]\rangle_{t}$ at time $t$ is understood
as averaging over the field probabilities
$p_{t}[\Phi ]$ as
\begin{equation}
\langle F[\Phi ]\rangle_{t} = \int {\cal D}\Phi\; p_{t}[\Phi ]\;
F[\Phi ].
\end{equation}
We stress that, in general, this ensemble averaging is not
thermal averaging since, out of equilibrium, we have no Boltzmann
distribution.
We shall only consider situations in which
\begin{equation}
\langle\rho_{j}({\bf r})\rangle_{t} = 0,
\end{equation}
That is,
we assume equal likelihood of a string or an antistring passing
through an infinitesimal area.  However, the line density
correlation functions
\begin{equation}
C_{ij}({\bf r} ;t) = \langle\rho_{i}({\bf r})\rho_{j}({\bf 0})\rangle_{t}
\end{equation}
will be non-zero, and give information on the persistence length of
strings.

It follows that, in terms of the zeroes of $\Phi ({\bf r})$,
$\rho_{i}({\bf r})$ can be written as
\begin{equation}
\rho_{i}({\bf r}) = \delta^{2}[\Phi ({\bf r})]\epsilon_{ijk}\partial_{j}
\Phi_{1}({\bf r}) \partial_{k}\Phi_{2}({\bf r}),
\label{rho}
\end{equation}
where
$\delta^{2}[\Phi ({\bf r})] = \delta[\Phi_{1} ({\bf r})] \delta[\Phi_{2}
({\bf r})]$.
The coefficient of the $\delta$-function in (\ref{rho}) is
the Jacobian of the transformation from line zeroes to field zeroes.
 It permits us to define a further line density that we shall also
find useful, the {\it total line density} $\bar{\rho}({\bf r})$
\begin{equation}
\bar{\rho_{i}}({\bf r}) = \delta^{2}[\Phi ({\bf r})]|\epsilon_{ijk}\partial_{j}
\Phi_{1}({\bf r}) \partial_{k}\Phi_{2}({\bf r})|.
\label{rhobar}
\end{equation}
Unlike the case for $\rho_{i}({\bf r})$
\begin{equation}
n(t) = \; \langle\bar{\rho_{i}}({\bf r})\rangle_{t} \; > 0
\end{equation}
and measures the {\it total} string density in the direction $i$, without
regard to string orientation.  The isotropy of the initial state
guarantees that $n(t)$ is independent of the direction $i$.
We note that the Jacobian factor multiplying the field
$\delta$-functions in (\ref{rho}) and (\ref{rhobar}) guarantees that
random field zeroes with no vorticity will not be counted.

In general, the best that we can do is write $\langle\bar{\rho_{i}}({\bf
r})\rangle_{t}$ as
\begin{eqnarray}
\langle\bar{\rho_{i}}({\bf
r})\rangle_{t} &=& \int {\cal D}\Phi\; p_{t}[\Phi ]\;
\delta^{2}[\Phi ({\bf r})]|\epsilon_{ijk}\partial_{j}
\Phi_{1}({\bf r}) \partial_{k}\Phi_{2}({\bf r})|
\nonumber
\\
&=& \int {\cal D}\alpha\,\,
\langle |\epsilon_{ijk}\partial_{j}
\Phi_{1}({\bf r}) \partial_{k}\Phi_{2}({\bf r})|\,
e^{i\int d{\bf x}\,\alpha_{a}\Phi_{a}}\rangle_{t}
\label{robart}
\end{eqnarray}
Our simple model assumes that $p_{t}[\Phi ({\bf r})]$ is Gaussian.
Details as to why this could be will be given later,
but if it is so then $n$ and
$C_{ij}$ are easily calculable.
Given that
\begin{equation}
\langle\Phi_{a}({\bf r})\rangle_{t} = 0,
\end{equation}
suppose that
\begin{equation}
\langle\Phi_{a}({\bf r})\partial_{j}\Phi_{b}({\bf r})\rangle_{t} = 0,
\end{equation}
that
\begin{eqnarray}
\langle\Phi_{a}({\bf r})\Phi_{b}({\bf r}')\rangle_{t} &=& W_{ab}(|{\bf r} -{\bf
r} '|;t)
\nonumber
\\
&=& \delta_{ab} W(|{\bf r} -{\bf r} '|;t),
\end{eqnarray}
and all other connected correlation functions are zero.

Then all ensemble averages are given in terms of $W(r;t)$, where $r
= |{\bf r}|$.  In particular, $\langle\bar{\rho_{i}}({\bf
r})\rangle_{t}$ separates as
\begin{equation}
\langle\bar{\rho_{i}}({\bf r})\rangle_{t} =
\langle \delta^{2}
[\Phi ({\bf r})]\rangle_{t}\;\langle |\epsilon_{ijk}\partial_{j}
\Phi_{1}({\bf r}) \partial_{k}\Phi_{2}({\bf r})|\rangle_{t}
\end{equation}
or, equivalently, from (\ref{robart})
\begin{equation}
\langle\bar{\rho_{i}}({\bf
r})\rangle_{t} = \int {\cal D}\alpha\,\,
\langle |\epsilon_{ijk}\partial_{j}
\Phi_{1}({\bf r}) \partial_{k}\Phi_{2}({\bf r})|\rangle_{t}\,\langle
e^{i\int d{\bf x}\,\alpha_{a}\Phi_{a}}\rangle_{t}.
\end{equation}
It follows \cite{halperin}, on first performing the $\alpha$ integration
in the second factor, that
\begin{equation}
n(t) =
\frac{1}{2\pi}\biggl |\frac{W''(0;t)}{W(0;t)}\biggr |,
\label{ni}
\end{equation}
where primes on $W$ denote differentiation with respect to r.
Thus
\begin{equation}
n(t) = O\biggl (\frac{1}{\xi^{2}}\bigg ),
\label{n}
\end{equation}
where $\xi$ is the length at time $t$ that sets the scale in $W(r;t)$.

On decomposing the density-density correlation
functions as
\begin{equation}
C_{ij}({\bf r} ;t) = A(r;t)\delta_{ij} +
B(r;t)\biggl(\frac{r_{i}r_{j}}{r^{2}} - \delta_{ij}\biggr),
\label{ddc}
\end{equation}
then, in the same Gaussian approximation,
 $A$ and $B$ can also be calculated in terms of $W$ and its
derivatives, with rather more difficulty, as
\begin{equation}
A = \frac{W'}{2\pi^{2}r\Delta^{2}}(W''\Delta + (W')^{2}W)
\label{defA}
\end{equation}
and
\begin{equation}
B = \frac{(W')^{2}}{2\pi^{2}r^{2}\Delta}
\end{equation}
in which
\begin{equation}
\Delta (r;t) =W(0;t)^{2} - W(r;t)^{2}.
\end{equation}
We note that both the density $n(t)$ and the correlation functions $C_{ij}({\bf
r}
;t)$ are independent of the scale of $W$.
The reader is referred to \cite{halperin} for further details.

To understand what $A(r;t)$ and $B(r;t)$ measure,
suppose ${\bf x} = (0,0,z)$.  Then $C_{ij}$ is diagonal, with
non-zero elements
\begin{equation}
C_{33} = A,\,\,  C_{11} = C_{22} = A - B
\end{equation}
For the sake of argument, consider the idealised situation in which there is
exactly {\it one}
string passing through each area $n(t)^{-1}$ in the $1-2$ plane and that these
areas
form a regular lattice of cell-length $\xi$.  Normalise $C_{ij}, A, B$ with
respect to $n^2$,
as
\begin{equation}
c_{ij} = \frac{C_{ij}}{n^2},\,\, a=\frac{A}{n^2},\,\, b =
\frac{B}{n^2},
\label{abc}
\end{equation}
whence $c_{11} = a-b$, etc.. The situation in which strings
passing through adjacent faces of this lattice are oppositely
oriented (i.e. string-antistring) then essentially corresponds to $c_{11}(\xi
;t) = c_{22}(\xi ;t) = -1$, a case of maximum anticorrelation.  On
the other hand, if there is equal likelihood of the next face containing
a string or an antistring then $c_{11}(\xi
;t) = c_{22}(\xi ;t) = 0$.  As for $C_{33}/n = c_{33}$, this takes
the value $c_{33}(\xi ;t) = 1$
if the string is guaranteed to continue in the same $3$-direction
through the next lattice cell, and takes the value $c_{33}(\xi ;t) = -1$ if
it changes direction at intervals $\xi$.
$B$ itself
can be isolated by observing that, if ${\bf x} =
\frac{1}{\sqrt{2}}(r,r,0)$, then $C_{12} = \frac{1}{2}B$. Thus a
value $b(\xi ;t) = 1$ is also a guarantee that the string changes
direction every $\xi$.  As we shall see, the situation is more
complicated, even in our simple model.
Nonetheless, we shall interpret strong {\it anticorrelation} in the
diagonal $c_{ii}$  (and positive $b$) on scales $\xi $ as signifying a
persistence
length (the typical length along the string before it has completely
changed direction) of $\xi$.

Nothing that we have said so far discriminates between vortices in a
relativistic or a non-relativistic medium.  This distinction appears
in the definition of $\langle ...\rangle_t$, to which we now turn.

\section{Field Dynamics with a Chemical Potential}

In order to compare vortex formation in relativistic and
non-relativistic media it is convenient to develop a single
formalism that encompasses both.  For simplicity we assume in each
case that the system is initially in {\it equilibrium} in the
disordered phase and that the
transition to the ordered phase occurs as the result of a 'quench', a rapid
change in the
environment of the system.  The interpolation between relativistic and
non-relativistic regimes is then
effected by introducing the
chemical potential $\mu$, coupled to the conserved charge $Q$
arising from the $O(2)$ symmetry.   Provided $\mu$ is small in
comparison to a particle mass the introduction of such
a potential will have little effect on a phase transition for the
relativistic theory initiated by quenching the system from a high
temperature $T_{0}$ to, effectively, zero temperature.
However, on increasing $\mu$ prior to the transition it becomes more costly to
produce
antiparticles and, if the initial temperature $T_{0}$ is decreased to a
value much less than $\mu$, antiparticles are frozen out.  The system is
then one of non-relativistic particles at a temperature much less than the
particle rest
masses. In this
nonrelativistic regime the transition is induced by a quench in
$\mu$ itself or equivalently, since $\mu$ determines the density, by
a density (pressure) quench.

\begin{figure}[htb]
\epsfxsize=8cm
\epsffile{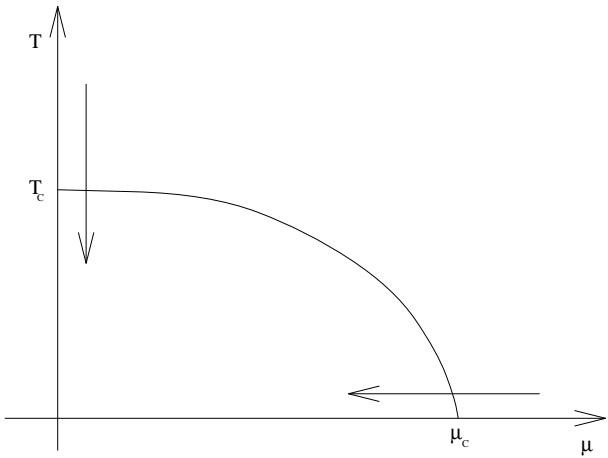}
\vspace{0.5cm}
\caption{Equilibrium phase structure of the global $O(2)$ theory.
The arrows show the directions in which the quenches are performed.}
\label{figphase}
\end{figure}

The situation is summarised in Fig.\ref{figphase}, which shows the
(equilibrium)
phase structure of the global $O(2)$ theory.
\typeout{Figure: EPS of phase structure}
The inner sector in
the $T-\mu$ quadrant is the ordered phase, the outer region the disordered
phase, and the line separating them the phase boundary.  $T_{c}$ is
the critical temperature at zero chemical potential  and
$\mu_{c}$ is the critical chemical potential at zero temperature.
The relativistic quench discussed in $I$ corresponds to a transition
at zero (small) $\mu$ across the phase boundary at $T_{c}$.  The
non-relativistic quench with which we shall be comparing it
corresponds to a transition at small $T$ across the phase boundary
near $\mu_{c}$.
\begin{figure}[htb]
\begin{center}
\setlength{\unitlength}{0.5pt}
\begin{picture}(495,200)(35,600)
\put(260,640){\makebox(0,0)[lb]{\large $C_-$}}
\put(185,750){\makebox(0,0)[lb]{\large $\Im m(t)$}}
\put(250,710){\makebox(0,0)[lb]{\large $C_+$}}
\put(510,645){\makebox(0,0)[lb]{\large $\Re e (t)$}}
\put(400,705){\makebox(0,0)[lb]{\large $t_f$}}
\put( 50,705){\makebox(0,0)[lb]{\large $t_0$}}
\thicklines
\put(400,690){\circle*{10}}
\put( 40,680){\vector( 1, 0){490}}
\put( 60,690){\vector( 1, 0){220}}
\put(280,690){\line( 1, 0){220}}
\put(500,680){\oval(10,20)[r]}
\put(500,670){\vector(-1, 0){220}}
\put(280,670){\line(-1, 0){220}}
\put(180,600){\vector( 0, 1){150}}
\end{picture}
\end{center}
\caption{The closed timepath contour $C_+ \oplus C_-$.}
\end{figure}
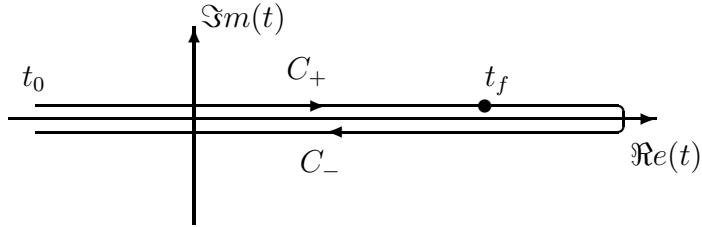
{}From this viewpoint the chemical potential is seen as determining
the initial conditions for the subsequent dynamics, for which
 we adopt the closed time path
method (Schwinger-Keldysh formalism)
\cite{schwinger,mahantapa,keldysh,Go4},
generalising the analysis begun in $I$ for the relativistic theory.
As a starting point suppose that, at the initial time $t_{0}$, we
are
 in a disordered state
with $\langle \phi \rangle = 0$.
Our ignorance is parametrised by the probability
distribution  $p_{t_0}[\Phi]$ that, at time $t_0$, $\phi(t_0,{\bf x}) =
\Phi({\bf x})$.
For the moment we take it as given.  Whether we are in a
relativistic or non-relativistic regime is largely encoded in $p_{t_0}[\Phi]$.
The subsequent, essentially generic,  non-equilibrium field evolution is driven
by a change
in the environment.  Specifically, for $t > t_{0}$ the action for the field is
taken to be
\begin{equation}
S[\phi] = \int d^{4}x \biggl (
\frac{1}{2} \partial_{\mu} \phi_a \partial^{\mu} \phi_a - \frac{1}{2}
m^{2}(t) \phi_a^2 - \frac{1}{4} \lambda (t) (\phi_a^2)^2
\biggr ).
\label{St}
\end{equation}
where $m(t)$, $\lambda (t)$ describe the evolution of the parameters
of the theory under external influences, to which the field responds.
As with $\Phi$, it is convenient to decompose $\phi$ in terms of two massive
real scalar fields $\phi_a$, $a=1, 2$ as $\phi = (\phi_1 + i\phi_2)/\sqrt{2}$,
in terms of which $S[\phi ]$ shows a global $O(2)$ invariance, broken by the
mass
term if $m^{2}(t)$ is negative.

The change of phase that begins at time $t_{0}$ will, by the
mechanism indicated earlier, lead to the appearence of vortices.
We saw in the previous section that the vortex distributions at later
times
 $t_f>t_0$ can be read
off from the probability $p_{t_f}[\Phi_f]$ that the
measurement of $\phi$ will give the value $\Phi_f$.  The evolution
of $p_{t}[\Phi ]$ from $t_{0}$ to $t_{f}$
is most simply
written as a closed time-path integral
in which the field $\phi$ is integrated along the closed path $C_+ \oplus C_-$
of Fig.2, where $\phi =\phi_+$ on $C_+$ and $\phi= \phi_-$ on $C_-$.

If ${\cal D} \phi_{\pm} = \prod_{a=1}^2 {\cal D} \phi_{\pm ,a}$ and spatial
labels are suppressed then
\begin{equation}
p_{t_f}[\Phi_f] = \int {\cal D} \Phi\, p_{t_0}[\Phi]
\int_{\phi_{\pm} (t_0) = \Phi} {\cal D} \phi_+  {\cal D}
\phi_- \, \delta [ \phi_+(t_f) - \Phi_f ] \, \exp \biggl \{ i \biggl (
S[\phi_+] - S[\phi_-] \biggr ) \biggr \}.
\end{equation}
where $\delta [ \phi_+(t) - \Phi_f ]$ is a delta functional, imposing
the constraint $\phi_+({\bf x},t) = \Phi_f ({\bf x})$ for each ${\bf x}$.
This is no more than the statement that, for a given initial state,
 the probability amplitude is
given by the integration along $C_{+}$, and its complex conjugate
(which, when multiplied with it, gives the probability) is given by
the integration back along $C_{-}$.
The $\pm$ two-field notation is misleading in that it suggests that the
$\phi_{+}$ ($=\phi_{a,+}$) and $\phi_{-}$ fields are decoupled.  That this is
not so
follows immediately from the fact that $\phi_{+}(t_f) =\phi_{-}(t_f)$.

To return to the initial conditions, we said that we could only achieve the
simple
analytic results of the previous section if $p_{t}[\Phi ]$ were
Gaussian and, therefore, that $p_{t_0}[\Phi]$ itself be
Gaussian.
The simplest such distribution has
$\Phi$  Boltzmann distributed at time $t_0$ at a
temperature of $T_0 = \beta_0^{-1}$ and chemical potential $\mu$
according to a {\it free}-field Hamiltonian
$H_0$,
which we take as
\begin{equation}
H_{0} = \int d^{3}x \biggl [ \frac{1}{2}\pi_{a}^{2} +
\frac{1}{2} (\nabla\phi_{a} )^{2} + \frac{1}{2} m^2 \phi_{a}^2
\biggr ].
\label{H_{0}}
\end{equation}
where $\pi_{a} = \dot{\phi_{a}}$.

We shall motivate the use of this free-field Hamiltonian in
specifying the initial field distribution later, when
we implement our Gaussian approximations.  However, since our
comments on extracting the non-relativistic limit from a
relativistic field theory have greater applicability than our particular
model we would like to be more general and extend $H_{0}$
to include interactions as
\begin{equation}
H_0 = \int d^{3}x \biggl [\frac{1}{2}\pi_{a}^{2}
+\frac{1}{2}(\nabla\phi_{a})^{2}
 + \frac{1}{2}m^2 \phi_{a}^{2} +
\frac{1}{4}\lambda(\phi_{a}^{2})^{2}\biggr ],
\label{H2}
\end{equation}
without enforcing a Gaussian straightjacket.
We stress that Gaussian does not necessarily mean {\it free}.
Most simply, particles need to interact before they can
equilibriate.  More importantly, the free-field Gaussian approximation adopted
in $I$ and here
can be extended to include interactions self-consistently
in a Hartree approximation \cite{boyanovsky}.  This will be considered
elsewhere.

Since chemical potentials are not usually relevant to
relativistic bosons a few words are in order.
The $O(2)$ invariance of $H_0$ leads to a conserved Noether current with
conserved charge
\begin{eqnarray}
Q &=& \int d^{3}x\; (\phi_{2}\pi_{1} - \pi_{2}\phi_{1})
\nonumber
\\
&=& \int d^{3}x\; (\phi_{2}\dot{\phi_{1}} - \dot{\phi_{2}}\phi_{1}).
\label{Q}
\end{eqnarray}
(in Minkowski space).  The numerical value of $Q$ is the number of particles
minus the
number of antiparticles
\footnote{For a relativistic theory the natural next
step would be to gauge the $O(2)$ or $U(1)$ symmetry by the
introduction of the electromagnetic field, whereby $Q$ becomes
proportional to the electric charge of the system.  With
non-relativistic media in mind, our constituents are neutral (e.g.
$He$ atoms) and we shall not do this.}.
The thermal probability distribution $p_{t_0}[\Phi]$ is thus taken
to be
\begin{equation}
p_{t_0}[\Phi] = \langle \Phi,t_0 | e^{- \beta_{0} (H_0 - \mu Q)} | \Phi,t_0
\rangle .
\end{equation}

There are two ways to proceed.  The first, which to us is the
most natural, accepts $H_0$ as
determining the temporal evolution of the physical fields, and relegates $\mu$
to the boundary conditions on the fields.
{}From this viewpoint, $p_{t_0}[\Phi]$ can be written as the
imaginary-time path integral
\begin{equation}
p_{t_0}[\Phi] =
\int_{B_{\mu}[\Phi]} {\cal D} \phi
\exp \biggl \{ i S_0 [\phi ] \biggr \},
\label{poo}
\end{equation}
for a corresponding action
\begin{equation}
S_{0}[\phi] = \int d^{4}x \biggl [
\frac{1}{2} (\partial_{\nu} \phi_{a} )(\partial^{\nu} \phi_{
a} ) - \frac{1}{2} m^2 \phi_{a}^2 - \frac{1}{4}\lambda(\phi_{a}^{2})^{2}
\biggr ].
\label{S_{0}}
\end{equation}
The label {\it zero} on $S_{0}[\phi]$ (and $H_0$ previously)
is a reminder that all are defined at
time $t_{0}$ (and not that the theory is free).  The
boundary condition $B_{\mu}[\Phi]$ incorporates the chemical potential.
In terms of the eigenstates of $Q$,
the complex field $\phi = (\phi_{1} +
i\phi_{2})/\sqrt{2}$ and its adjoint, $B_{\mu}$ becomes
\begin{equation}
B_{\mu}[\Phi ]:\,\phi (t_0) = \Phi =
e^{-\beta_{0}\mu q}\phi (t_0-i \beta_0),
\end{equation}
where $q = 1$ is the $\phi$-field  eigenvalue of $Q$ and the
time-integral in (\ref{S_{0}}) is taken in imaginary time from
$t_0$ to $t_0 -i\beta_{0}$.  For $\phi^{*}$, $q = -1$.
In terms of
the same complex fields we have
\begin{equation}
S_{0}[\phi] = \int d^{3}x [(\partial_{\nu} \phi^{*})(\partial^{\nu}\phi ) - m^2
\phi^{*}\phi - \lambda(\phi^{*}\phi)^{2}]
\end{equation}
and, with $\pi = (\pi_{1} - i\pi_{2})/\sqrt{2}$,
\begin{equation}
H_{0} = \int d^{3}x [ \pi^{*}\pi +(\nabla\phi^{*} )(\nabla\phi )
 + m^2 \phi^{*}\phi + \lambda(\phi^{*}\phi)^{2}]
\label{Hc2}
\end{equation}
However, for calculational purposes
a decomposition in terms of $\phi_{a}, \pi_{a}$ is usually preferable.

Although we are just setting an initial condition the effect is, inevitably, to
give an
action $S_{0}[\phi]$ of the form of $S[\phi]$ of (\ref{St}).
This permits the interpretation that the action  $S[\phi]$ is
valid for all times $t$, with the proviso that the system is in thermal
equilibrium
for  $t<t_0$, during
which period the mass $m(t)$ takes the constant value $m$ and $\lambda(t) =
\lambda$, also constant.

On relabelling the integration variable $\phi$ of (\ref{poo}) by $\phi_3$,
we now have the explicit form for $p_{t_f}[\Phi_f]$:-
\begin{eqnarray}
\nonumber
p_{t_f}[\Phi_f] &=& \int {\cal D} \Phi \int_{B_{\mu}[\Phi]}
{\cal D} \phi_3 \,  e^{i S_0[\phi_3]}
\int_{\phi_{\pm}(t_0) = \Phi}
 {\cal D} \phi_+ \Phi {\cal D} \phi_- \,
e^{i ( S[\phi_+] - S[\phi_-] ) } \delta [ \phi_+(t_f) - \Phi_f ]
\\
\nonumber
&=& \int_{B_{\mu}} {\cal D} \phi_3 {\cal D} \phi_+ {\cal D} \phi_- \, \exp
\biggl \{
i S_0[\phi_3] + i ( S[\phi_+] - S[\phi_-] )
\biggr \} \,
\delta [ \phi_+(t_f) - \Phi_f ],
\end{eqnarray}
where the boundary condition $B_{\mu}$ is now (in terms of the field
combinations $\phi = (\phi_{1} +
i\phi_{2})/\sqrt{2}$)
\begin{equation}
B_{\mu}:\,\,\phi_{+}(t_0) =
e^{-\beta_{0}\mu q}\phi_3(t_0- i \beta_0).
\end{equation}
More succinctly, $p_{t_f}[\Phi_f]$ can be written as the time ordering of a
single field doublet:-
\begin{equation}
p_{t_f} [ \Phi_f] = \int_{B_{\mu}} {\cal D} \phi \, e^{i S_C [\phi]} \, \delta
[
\phi_+ (t_f) - \Phi_f ],
\label{pf}
\end{equation}
along the contour $C=C_+ \oplus C_- \oplus C_3$ of Fig.3, extended to include a
third imaginary leg, where $\phi$ takes the values $\phi_+$, $\phi_-$
and $\phi_3$ on $C_+$, $C_-$ and $C_3$ respectively, for which $S_C$
is $S[\phi_+]$, $S[\phi_-]$ and $S[\phi_3]$, for which last case
$m(t) = m, \lambda (t) = \lambda$, $t\in C_{3}$.
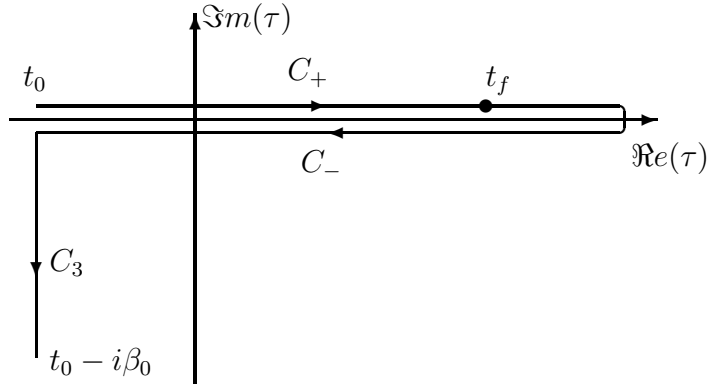
\begin{figure}[htb]
\begin{center}
\setlength{\unitlength}{0.5pt}
\begin{picture}(495,280)(35,480)
\put( 70,565){\makebox(0,0)[lb]{\large $C_3$}}
\put(260,640){\makebox(0,0)[lb]{\large $C_-$}}
\put(185,750){\makebox(0,0)[lb]{\large $\Im m(\tau)$}}
\put(250,710){\makebox(0,0)[lb]{\large $C_+$}}
\put(510,645){\makebox(0,0)[lb]{\large $\Re e (\tau)$}}
\put(400,705){\makebox(0,0)[lb]{\large $t_f$}}
\put( 50,705){\makebox(0,0)[lb]{\large $t_0$}}
\put(70,490){\makebox(0,0)[lb]{\large $t_0-i \beta_0$}}
\thicklines
\put(400,690){\circle*{10}}
\put( 40,680){\vector( 1, 0){490}}
\put( 60,690){\vector( 1, 0){220}}
\put(280,690){\line( 1, 0){220}}
\put(500,680){\oval(10,20)[r]}
\put(500,670){\vector(-1, 0){220}}
\put(280,670){\line(-1, 0){220}}
\put( 60,670){\vector( 0,-1){110}}
\put( 60,560){\line( 0,-1){ 60}}
\put(180,480){\vector( 0, 1){280}}
\end{picture}
\end{center}
\caption{A third imaginary leg}
\end{figure}

As a final step in these formal manipulations we see that expression
(\ref{pf}) enables us to write the $\Phi$-field ensemble averages
$\langle ...\rangle_{t}$ in terms of the $\phi$-field thermal Wightman
functions.
Consider the generating
functional:-
\begin{equation}
Z_{\mu}[j_+,j_-,j_3] = \int_{B_{\mu}} {\cal D} \phi \, \exp \biggl \{ i
S_C[\phi] +
i \int j \phi \biggr \},
\label{Z}
\end{equation}
where $\int j \phi$ is a short notation for:-
\begin{equation}
\int j \phi \equiv \int_0^{\infty} dt \, \,  [ \, j_+(t) \phi_+(t) - j_-
\phi_-(t) \, ] \, + \int_0^{-i \beta} j_3(t) \phi_3(t) \, dt,
\end{equation}
omitting spatial arguments. Then introducing $\alpha_a({\bf x})$ where
$a=1, 2$, we find:-
\begin{eqnarray}
p_{t_f} [\Phi] &=& \int {\cal D} \alpha \int_{B_{\mu}} {\cal D } \phi \,\,
\exp \biggl \{i
S_C[\phi] \biggr \} \,\,
\exp \biggl \{ i\int d^{3} x \alpha_a({\bf x}) [ \phi_+(t_f,{\bf
x}) - \Phi({\bf x}) ]_a \biggr \}
\nonumber
\\
&=& \int {\cal D} \alpha \, \, \exp \biggl \{ -i \int \alpha_a \Phi_a
\biggr \} \,Z_{\mu}[\overline{\alpha},0,0],
\end{eqnarray}
where $\overline{\alpha}$ is the source $\overline{\alpha}({\bf x},t) =
\alpha({\bf x}) \delta (t-t_f)$. As with ${\cal D} \phi$, ${\cal D}
\alpha$ denotes $\prod_1^N {\cal D} \alpha_a$.

Ensemble averages are now expressible in terms of $Z_\mu$.  Of
particular relevance,
$W_{ab}(|{\bf r} -{\bf r} '|;t) = \langle\Phi_{a}({\bf r})\Phi_{b}({\bf
r}')\rangle_{t}$ is given by
\begin{eqnarray}
W_{ab}(|{\bf r} -{\bf r} '|;t) &=& \int {\cal D}\Phi\,\, \Phi_{a}({\bf
r})\Phi_{b}({\bf
r}')
\int {\cal D} \alpha \,\,\exp \biggl \{ -i \int \alpha_a \Phi_a
\biggr \} \,\,Z_{\mu}[\overline{\alpha},0,0]
\nonumber
\\
&=&
-\int {\cal D}\alpha\,\, \frac{\delta^{2}}{\delta\alpha_{a}({\bf
r})\delta\alpha_{b}({\bf
r}')}
\int {\cal D} \Phi\,\,\exp \biggl \{ -i \int \alpha_a \Phi_a
\biggr \}\,\, Z_{\mu}[\overline{\alpha},0,0]
\nonumber
\\
&=&
-\int {\cal D}\alpha\,\,\frac{\delta^{2}}{\delta\alpha_{a}({\bf
r})\delta\alpha_{b}({\bf
r}')}
\biggl \{\delta^{2}\,[ \alpha ]\,\,Z_{\mu}[\overline{\alpha},0,0]\biggr \}
\end{eqnarray}
On integrating by parts
\begin{eqnarray}
W_{ab}(|{\bf r} -{\bf r} '|;t) &=&
-\frac{\delta^{2}}{\delta\alpha_{a}({\bf r})\delta\alpha_{b}({\bf
r}')}Z_{\mu}[\overline{\alpha},0,0]\bigg |_{\alpha = 0}
\nonumber
\\
&=& \langle\phi_{a}({\bf r},t)\phi_{b}({\bf
r}',t)\rangle,
\end{eqnarray}
the equal-time thermal Wightman function with the given thermal boundary
conditions.
Because of the time evolution
there is no time translation invariance in the double time label.

Not surprisingly, $p_{t} [\Phi ]$ can only be calculated explicitly in
very simple circumstances, like our Gaussian approximation, but
before we do that we still have to extract the relativistic limit.
Since the chemical potential is embedded in the equilibrium boundary
conditions and, like temperature, can only be defined in
equilibrium, this is essentially an equilibrium problem and, for
the moment,we forget the dynamics.

\section{Manipulating the Chemical Potential}

The expression (\ref{pf}) is valid for all $\mu$ and all $T_{0}$
but, as it stands, is not
sympathetic to the isolation of a non-relativistic limit.
Nonetheless, the extraction of a non-relativistic regime from it is
not difficult.
Instead of working with $p_{t_{0}}[\Phi ]$ directly, we can work
with the partition function $Z_{\mu}$, as we saw in the previous section.
The partition function for this equilibrium theory is (for doublet
sources $j_{a}$ restricted to the $C_{3}$-contour)
$Z_{\mu}[0,0,j_{3}]$ of (\ref{Z}), written
\begin{equation}
Z_{\mu}[j]
\equiv Z_{\mu}[0,0,j]
= \int_{B_{\mu}} {\cal D} \phi \, \exp \biggl \{ i S_{0}[\phi] +
i \int j_{a} \phi_{a} \biggr \},
\label{Z_{0}}
\end{equation}
where $S_{0}[\phi]$ is given in (\ref{S_{0}}) and we integrate only
along the contour $C_{3}$.
  As before, ${\cal D}\phi\equiv{\cal
D}\phi_{1}{\cal D}\phi_{2}$.  On rotating to Euclidean time $\tau =
it$, $Z_{\mu}[j]$ becomes
 \begin{equation}
Z_{\mu}[j] = \langle\exp \biggl \{\int j_{a} \phi_{a} \biggr \}\rangle =
 \int_{B_{\mu}} {\cal D} \phi \, \exp \biggl \{ - S_{0,E}[\phi] -
\int j_{a} \phi_{a} \biggr \},
\label{Z_{E}}
\end{equation}
where $S_{0,E}[\phi]$ is the relativistic Euclidean action
\begin{equation}
S_{0,E} = \int_{0}^{\beta_{0}} d\tau \int d^{3}x \biggl [
\frac{1}{2}\dot{\phi}_{a}^{2} +
\frac{1}{2} (\nabla\phi_{a} )^{2} + \frac{1}{2} m^2 \phi_{a}^2
- \frac{1}{4}\lambda(\phi_{a}^{2})^{2}
\biggr ].
\label{S_{0,E}}
\end{equation}
and the sum is taken over fields $\phi = (\phi_1 +
i\phi_2)/\sqrt{2}$
 satisfying the boundary condition $B_{\mu}$: $\phi ({\bf
x}, \tau) = e^{-\beta_{0}\mu}\phi ({\bf
x}, \tau-\beta_{0})$ in imaginary
time.  The dot now means differentiation with
respect to $\tau$, $\dot{\phi_{a}}= \partial_{\tau}\phi_{a}$.

The bracket $\langle ...\rangle$ here denotes the thermal average
\begin{equation}
\langle F[\phi]\rangle = Tr\{e^{-\beta_{0} (H_{0} - \mu Q)}\; F[\phi]\}.
\label{chemav}
\end{equation}
where $H_{0}$ is given in (\ref{H2}).
Because of its time-independence we have dropped the $t$-suffix.
The net charge $\bar{Q}$ (excess of particles over antiparticles) in the
system is obtained by differentiating $Z_{\mu}[j] = \langle e^{\int
j\phi}\rangle$ as
\begin{equation}
\bar{Q} = \langle Q\rangle =
\frac{1}{\beta_{0}}\frac{\partial}{\partial\mu}\; \ln Z_{\mu}[j]\bigg|_{j=0}.
\end{equation}
This determines $\mu$ in terms of the chosen $\bar{Q}$ and $T_{0}$.
In a non-relativistic environment in which antiparticles are not
present, $\bar{Q}$ becomes the mean particle number (and hence is
proportional to the density).

$Z_{\mu}[j]$  suffers from
the
presence of the chemical potential in the boundary conditions, which
makes it difficult to identify the phase of the system readily.
This is
clarified by adopting a second approach to chemical potentials, in
which we transfer the chemical potential term $\mu
Q$ to the
Hamiltonian, to create an effective Hamiltonian $H_0 - \mu Q$.
$H_0 - \mu Q$ no longer generates time translations of the
physical fields $\phi_{a}$ or, equivalently, $\phi (t)$ and
$\phi^{*}(t)$.  However, it does generate time translations of the effective
fields $\tilde{\phi}(t) = e^{i\mu qt}\phi (t) = e^{i\mu t}\phi (t)$
and $\tilde{\phi^{*}}(t)$, with $q = -1$.
It follows that $\tilde{\phi}(t)$ and $\tilde{\phi}^{*}(t)$ integrations, and
hence
$\tilde{\phi}_{a}$
integrations, are now taken
over {\it periodic} configurations with boundary conditions
$B_{0}:\tilde{\phi}_{a} ({\bf
x}, \tau) = \tilde{\phi}_{a}({\bf
x}, \tau-\beta_{0})$ in imaginary
time, with no $\mu$-dependence
Details are given in the Appendix.

The advantages of removing $\mu$ from the boundary conditions become
apparent when we express $Z_{\mu}$ of
(\ref{Z_{0}}) not as a sum over $\phi$-field
histories but as a
sum over histories in the $\tilde{\phi}$-fields.
Although not usually posed this way, the details are
well-understood. For
example, see Kapusta \cite{kapusta}, Haber $\&$ Weldon \cite{art},  and
Bernstein
et al.
\cite{dodelson}.
On direct substitution, $Z_{\mu}[j]$ takes the form
\begin{equation}
Z_{\mu}[j] =
 \int_{B_{0}} {\cal D} \tilde{\phi} \, \exp \biggl \{ - S_{0}[\tilde{\phi}
;\mu] -
\int j_{a} \tilde{\phi}_{a} \biggr \},
\label{Zmu}
\end{equation}
where $S_{0}[\phi ;\mu]$ is now the effective action
\begin{equation}
S_{0}[\phi ;\mu] = \int_{0}^{\beta_{0}} d\tau \int d^{3}x \biggl [
\frac{1}{2}\dot{\phi}_{a}^{2} +
\frac{1}{2} (\nabla\phi_{a} )^{2} + \frac{1}{2}m_{0}^2 \phi_{a}^2
+ i\mu (\phi_{2}\dot{\phi_{1}} - \dot{\phi_{2}}\phi_{1})
-\frac{1}{4}\lambda(\phi_{a}^{2})^{2}
\biggr ].
\label{S0}
\end{equation}
where
\begin{equation}
m_{0}^{2} = m^2-\mu^{2}
\end{equation}
and, as above, the sum is taken over periodic field configurations $B_{0}$,
rather than the boundary condition $B_{\mu}$.
Details are given in Kapusta \cite{kapusta} and \cite{dodelson}.
We note that $\phi_{a}$ and $\tilde{\phi}_{a}$ have the same zeroes
and hence are equally good for the calculation of vortices.
It is a separate exercise to rewrite $Z_{\mu}[j]$ in terms of the
conventional actions for a non-relativistic order field $\psi(t)$.
Details will be given elsewhere \cite{altimray2} but, in the
interim, the basic idea is given in recent conference proceedings
\cite{ray2,ray3} by
one of us (R.J.R).

This displacement of $\mu$ from the boundary conditions to the
action has enabled us to replace the classical potential by an
effective potential in which the state of the system is more transparent.
The relevant quantity is $m_{0}^{2}$, rather than
$m^2$ and $\mu^2$ separately
\footnote{We shall see later that,
in our approximations, the other term depending on $\mu$,
$i\mu (\phi_{2}\dot{\phi_{1}} - \dot{\phi_{2}}\phi_{1})$, plays no role.}.
Semiclassically, when $m_{0}^{2} > 0$ (i.e. $\mu^{2} < m^{2}$) the $O(2)$
symmetry is unbroken.
This describes our disordered initial state.
However, once $m_{0}^{2} < 0$ (i.e. $\mu^{2} > m^{2}$) the free theory is
unstable.
This is a signal that the $O(2)$ symmetry is broken and a transition
 to an ordered phase has occurred.  Thus, effectively, it is
$m_{0}^{2}$, rather than $m^2$ that carries the time dependence,
changing from positive to negative at $t = t_{0}$.

Let us first consider the relativistic regime in which the initial
environment is very hot, with $T\gg m\gg
\mu$.
In this case, with $\mu$ irrelevant, the symmetry is broken by a change in
$m^{2}$.
Suppose that, on the completion of the transition, $m_{0}^{2}\simeq
m^{2}$ takes the value $m_{0}^{2} = -M^{2} < 0$.
If, in this relativistic case, the final temperature is very low, there are
then no thermal effects
and $M$ is a physical parameter, determining the Higgs mass, $m_{H}
= \sqrt{2} M$.
In order to recover the phase boundary in Fig.1 in this case it is
necessary to include thermal radiative effects.
As it stands $m_{0}(t\leq 0) = m_{0}$
is not a physical parameter, but there is no loss in taking it as the effective
scalar field mass
at temperature $T_{0} = \beta_{0}^{-1}$.  That is,
in the mean-field approximation we take
\begin{equation}
m_{0}^{2} = -M^{2}\biggl ( 1-\frac{T_{0}^{2}}{T_{c}^{2}}\biggr )
\label{m00}
\end{equation}
where $T_{0}$ is greater than the transition temperature $T_{c}$,
given by $T_{c}^{2} = 3M^{2}/\lambda$ in the same
approximation.  It might be objected that this is inconsistent in that
this mass is defined in terms of fluctuations at scales
much larger than the typical domain size.  Since results will turn
out to be largely independent of $m_{0}$ provided it is comparable to
$M$ this is not really a
problem.  With the parametrisation above
this would not be true only if we quenched from very close to the
transition, and we do not consider this possibility yet.

On the other hand, for the non-relativistic theory, for which
$T/m\ll 1$, the parameter $m$ can always be identified with the
boson mass.  We
change the sign of $m_{0}^{2}$ in the
action by a change in $\mu$ from
$\mu^{2} < m^{2}$ to $\mu^{2} > m^{2}$, a density quench.
Equivalently, in terms of the non-relativistic chemical potential
\begin{equation}
\mu_{nr} = \mu - m
\end{equation}
we go from negative $\mu_{nr}$ to positive $\mu_{nr}$.
To preserve uniformity of notation, we take $m_{0}^{2}(t) =
m_{0}^{2}$ when $t\leq t_{0}$ and $m_{0}^{2}(t) = -M^{2} < 0$ once
the transition is complete.

\section{Vortex Formation}

Having established the role of the chemical potential in letting us
interpolate between relativistic and non-relativistic regimes in the
initial conditions, we are now in a position to determine the effect
of these initial conditions on vortex production in a simple model of
Gaussian field fluctuations.

We begin by recapitulating the methods adopted in $I$
for a relativistic regime, for which $\mu =0$.
We have already assumed that the initial conditions correspond
to a disordered state in equilibrium for $t < t_{0}$.  Specifically,
in $I$ we adopted an initial Gaussian distribution of fields arising
from the {\it quadratic}  Hamiltonian $H_0$ of (\ref{H_{0}}).
Although quadratic, the choice of $m_{0} > 0$ of (\ref{m00}) has thermal
interactions encoded within it.  We assume that interactions are
sufficiently weak that, having established equilibrium, their
effect on the Gaussian distribution is small.

For $t > t_0$, the system is forced to change.
The simplest assumption, made in $I$, was that, for $t > t_{0}$, $m^{2}(t)$
and $\lambda (t)$ could be taken as {\it constant}, an idealised
quench in which $m^{2}(t)$ changes sign to
take the {\it negative} value $m^{2}(t) = - M^2 <0$ {\it immediately}.  That
is,
the potential at the origin has been instantaneously inverted
everywhere, breaking the global $O(2)$ symmetry.  Given the initial
high temperature $T_0$, this can be thought of as a temperature
quench.  If
$\lambda (t) =  \lambda$ is very {\it weak} then, for times $Mt <
\ln(1/\lambda)$, the $\phi$-field, falling down the hill away from
the metastable vacuum, will not yet have experienced the upturn of
the potential, before the point of inflection at $\langle |\phi |\rangle =
\sqrt{2M^{2}/ 3\lambda}$.  Thus, for these small times,  $\lambda
(t)$ can also be set to zero, and $p_{t}[\Phi ]$ is Gaussian, as required.

The onset of the phase transition at time $t=t_0$ is characterised by
the instabilities of long wavelength fluctuations permitting the growth of
correlations. Although the initial
value of $\langle \phi \rangle$ over any volume is zero, we
anticipate that
the resulting phase separation or spinodal decomposition will lead to
domains of constant $\langle \phi \rangle$ phase, whose boundaries will
trap vortices.

We now turn to the case of non-relativistic vortex formation, cast
in as similar a way as possible.
The situation we have in mind is the following.  Consider a gas of
non-relativistic bosons of mass $m$ in thermal equilibrium at
temperature $T_{0}\ll m$.  The state is
disordered.

By changing the density we then force $\mu_{{\rm nr}}$ to change sign, from
$\mu_{0} < 0$ to
$\mu_{{\rm nr}} =  \mu_{f} > 0$, as rapidly as possible. Yet again,
in this idealised model
we suppose that this change is
implemented instantaneously everywhere at time $t = t_{0}$. This sets in
motion a change to a superfluid phase, in which the $O(2)$ particle symmetry
is broken by the condensate.  Order is again established by the
 growth of long wavelength fluctuations in
which domains form.  These domains
will trap vortices on their boundaries.

So as to give a solvable Gaussian theory, for which we can use the
results of Section 2, we maintain our Procrustean approach of $I$
and ignore interactions before and after the change in chemical
potential.    Our
initial condition is now equivalent to the statement that, for $t
\leq t_0$, the particles are distributed according to the Boltzmann
distribution
\begin{equation}
n(\epsilon ) = \frac{1}{e^{\beta_{0}(\epsilon -\mu_{0})} - 1},
\label{nnrel}
\end{equation}
where $\epsilon = k^{2}/2m$ and $\mu_{0} < 0$ is the
non-relativistic chemical potential $\mu - m$.

In the previous section we showed how the model
is that of the relativistic theory of the field $\phi$
with effective action $S_{0}[\phi ;\mu]$  of (\ref{S0}),
in which $|\mu_{nr}|\ll m$.
In the notation of that section, when $\mu\simeq m$, $m_{0}^{2}$
changes from
\begin{equation}
m_{0}^{2}(t) = m_{0}^{2} = -2m\mu_{0}\, > 0,
\end{equation}
when $t\leq t_{0}$, to
\begin{equation}
m_{0}^{2}(t)  = -M^{2} = -2m\mu_{f}\, < 0,
\label{ms}
\end{equation}
when $t > t_0$.
Henceforth, we take $t_{0} = 0$.

Of course, instantaneous change is physically impossible.
Consider small amplitude fluctuations of $\phi_a$, at the top of the
parabolic potential hill.  Long wavelength fluctuations, for which $|{\bf
k}|^2 < M^2$, begin to grow exponentially. If their growth rate
$\Omega (k) = \sqrt{M^2 - |{\bf k}|^2}$ is much slower than
the rate of change of the environment which is causing the quench,
then those long wavelength modes are unable to track the quench.
 It will
turn out that the time-scale at which domains appear in this
instantaneous quench is $t_d = O(M^{-1})$. As long as the time taken
to implement the quench is comparable to $t_d$ and much less than
$t_f = O(M^{-1}ln (1/\lambda )$ the approximation is relevant.

We note that, in the non-relativistic regime, $\Omega (k)$ has the same
definition,
but can be rewritten as
\begin{eqnarray}
\Omega^2(k) &=& M^2 - |{\bf k}|^2 = 2m\biggl (\mu_{f} -
\frac{k^{2}}{2m}\biggr )
\nonumber
\\
&=& 2m(\mu_{f} - \epsilon (k)).
\end{eqnarray}
Thus the momentum restriction $|{\bf k}|<M$ is just
$\epsilon (k) < \mu_{f}$.

We are now in a position to evaluate $p_t[\Phi]$, or rather $W_{ab}(r;t)$, for
$t > 0$, and
calculate the defect density accordingly.  Details are given in the Appendix.
Before we quote the result we note that
the $i\mu (\phi_{2}\dot{\phi_{1}} - \dot{\phi_{2}}\phi_{1})$
term in $S_{0}[\phi ;\mu]$ of (\ref{S0}) couples the $a =1$ and $a =
2$ fields $\phi_{a}$ together and, in general,
$G_{ab}({\bf
r} -{\bf r}';t,t') = \langle\phi_{a}({\bf r},t)\phi_{a}({\bf
r}',t')\rangle $
is {\it not} diagonal in the $O(2)$ labels.
However, for {\it equal} times diagonal behaviour is restored as
$G_{ab}({\bf r};t,t) = \delta_{ab}G({\bf r};t,t)$ and the
$i\mu (\phi_{2}\dot{\phi_{1}} - \dot{\phi_{2}}\phi_{1})$
term can effectively be discarded.  This leaves us in a situation
very like the original relativistic case of $I$ and Section 3 for
which the results of Halperin are directly applicable.  That is, $W_{ab}$ is
diagonal,
\begin{equation}
W_{ab}(|{\bf r}-{\bf r}'|;t,t) = \delta_{ab}W(|{\bf r}-{\bf r}'|;t,t),
\end{equation}
whence $W(|{\bf
r} -{\bf r}'|;t) = \langle\phi_{a}({\bf r},t)\phi_{a}({\bf
r}',t)\rangle $ (no summation), the
thermal Wightman function for either $\phi_{1}$ or $\phi_{2}$.

For all the utility of bringing the effective potential into the
action from the boundary conditions, $W(r;t)$ is still most easily
 built from the modes ${\cal
U}^{\pm}_{a,k}$, satisfying the equations of
motion
\begin{equation}
\biggl [ \frac{d^2}{dt^2} + {\bf k}^2 + m^2(t) \biggr ] {\cal
U}^{\pm}_{a,k} =0,
\end{equation}
for $m^{2}(t)$ above,
 subject to the initial condition of a thermal distribution with
chemical potential $\mu$
See the Appendix for
greater detail.

But for the chemical potential,
this situation of inverted harmonic oscillators was studied many
years ago by Guth and Pi \cite{guth} and Weinberg and Wu
\cite{weinberg}.  In the context of domain formation, we refer to
the recent work of Boyanovsky et al.,
\cite{boyanovsky}.
For the case in hand, if we make a separation
into the unstable long wavelength modes, for which $|{\bf k}|<M$,
and the short wavelength modes $|{\bf k}|>M$, then  $W(r;t)$
is the real quantity
\begin{eqnarray}
 W(r;t) &=& \int_{|{\bf k}|<M} d \! \! \! / ^3 k
\, e^{i {\bf k} . {\bf x} } C(k;\mu )
\biggl [ 1+ A(k)(\cosh(2\Omega (k)t) - 1 ) \biggr ]
\nonumber
\\
&+& \int_{|{\bf k}|>M} d \! \! \! / ^3 k
\, e^{i {\bf k} . {\bf x} } C(k;\mu )
\biggl [ 1+ a(k)(\cos(2w(k)t) - 1 ) \biggr ]
\label{G}
\end{eqnarray}
with $r = |{\bf x}|$ and
\begin{eqnarray}
\Omega^2(k) &=& M^2 - |{\bf k}|^2
\nonumber
\\
w^{2}(k) &=& -M^2 + |{\bf k}|^2
\nonumber
\\
A(k) &=& \frac{1}{2} \biggl (
1+ \frac{\omega ^2(k) }{\Omega^2(k)} \biggr )
\nonumber
\\
a(k) &=& \frac{1}{2} \biggl (
1- \frac{\omega ^2(k) }{w^2(k)} \biggr )
\label{defs}
\end{eqnarray}
All the $\mu$-dependence is contained in the factor
\begin{equation}
{\cal C}(k, \mu_{0}) = \frac{1}{2\omega (k)}\biggl [\frac{e^{-\beta_{0}\omega}
-e^{\beta_{0}\omega}}
{e^{\beta_{0} (m + \mu_{0})}  - e^{-\beta_{0}\omega} - e^{\beta_{0}\omega}
+ e^{-\beta_{0} (m + \mu_{0})}}.
\biggr ]
\end{equation}
where
\begin{equation}
\omega ^2(k) = |{\bf k}|^2 + m_0^2
\end{equation}
In the zero chemical potential limit $\mu = m_{0} + \mu_{0} = 0$, ${\cal C}(k,
\mu_{0})$ takes the familiar form
\begin{equation}
{\cal C}(k, \mu_{0}) = \frac{1}{2\omega (k)}coth(\beta_{0}\omega (k)/2)
\label{cr}
\end{equation}
On the other hand, in the non-relativistic limit $\epsilon (k) = {\bf k}^{2}/2m
\simeq \omega
- m$ and $\mu_{0}\ll m$ ${\cal C}(k, \mu_{0})$ is the equally
familiar Bose distribution
\begin{equation}
{\cal C}(k, \mu_{0}) \simeq  \frac{1}{2\omega (k)}\bigg (\frac{1}{1 -
e^{-\beta_{0}(\epsilon
(k) - \mu_{0})}}\Bigg )
\label{C}
\end{equation}
Further, in the high temperature relativistic limit $T_{0}\gg m_{0}$,
${\cal C}(k, \mu_{0})$ of (\ref{cr}) simplifies as
\begin{equation}
\frac{1}{2\omega (k)}coth(\beta_{0}\omega (k)/2)
\simeq \frac{T_{0}}{{\bf k}^{2} + m_{0}^{2}}.
\label{wc}
\end{equation}
Equally, in the non-relativistic limit when $\epsilon (k) - \mu_{0} \ll
T_{0}$,
\begin{equation}
{\cal C}(k, \mu_{0})
\simeq  \frac{T_{0}}{\epsilon (k) + |\mu_{0}|}
\label{C2}
\end{equation}
from (\ref{C}).
This reproduces (\ref{wc}),
\begin{equation}
{\cal C}(k, \mu_{0})\simeq
\frac{T_{0}}{{\bf k}^{2} + m_{0}^{2}}
\label{wC}
\end{equation}
up to an irrelavant coefficient of proportionality,
on using the definition of $m_{0}$ given earler in (\ref{ms}).
That is, in these regimes the relativistic and non-relativistic
$W(r,t)$ are
{\it identical},
once we take the identifications of (\ref{ms}) into account.

We observe that, if we were to use $W(r;t)$ of (\ref{G}) as it
stands, then both $W(0;t)$ and $W''(0;t)$ necessarily suffer from
ultraviolet divergences.
However, the string thickness at the end of the quench will be $O(M^{-1})$.
It is the zeroes  coarse-grained to this scale that will provide the
subsequent network.  Thus, if we do not probe the field zeroes
within a string we need consider only the first term
in (\ref{G}).  There is another point. Even before the quench begins
there is a high density of line zeroes coarse-grained to this same
scale $O(M^{-1})$ in the initial equilibrium phase.  From (\ref{n})) their
density is
\footnote{This was essentially the basis for Halperin's results on
string densities in \cite{halperin}.} $n(t) = O(M^2)$ if
$M\geq m$, which we assume.
However, these modes are entirely transient due to the uncertainty principle.
If we were to calculate the correlations of $\rho_{i}(\bf{x})$ at
{\it different} times $t$ and $t'$, we would find rapidly
oscillating behaviuor with period $\Delta t = O(m^{-1})$.  On the
other hand, a calculation of the density correlations at different
times from the unstable modes in (\ref{G}) does not give
oscillatory, but damped, behaviour.  It is the residue of the
strings produced by the unstable modes that survives to produce the
network, and the transient strings can be ignored.  Henceforth, we
retain only the first term
\begin{equation}
 W(r;t) = \int_{|{\bf k}|<M} d \! \! \! / ^3 k
\, e^{i {\bf k} . {\bf x} }{\cal C}(k, \mu_{0})
\biggl [ 1+ A(k)(\cosh(2\Omega (k)t) - 1 ) \biggr ]
\label{Gl}
\end{equation}
of (\ref{G}).  In the two critical regimes where (\ref{wc}) and
(\ref{wC}) are valid, $W(r;t)$ is again the same for both cases.

Even though the approximation is only
valid for small times, there is a regime $Mt\ge 1$, for small couplings, in
which
$t$ is large enough for $\cosh(2Mt) \approx \frac{1}{2} \exp(2M
t)$ and yet $Mt$ is still smaller than the time $O(\ln 1/\lambda)$ at
which the fluctuations begin to sample the ground-state manifold.
In this regime
\begin{equation}
W(r;t)\simeq \int_{|{\bf k}|<M} d \! \! \! / ^3 k
\, {\cal C}(k, \mu)
e^{i {\bf k} . {\bf x} }
A(k)\;e^{2\Omega (k)t}
\end{equation}
In
these circumstances the integral at time t is dominated by a
peak in the integrand $k^{2} e^{2\Omega (k)t}$ at $k$ around $k_c$, where
\begin{equation}
t k_c^2 = M \biggl ( 1 + O \biggl ( \frac {1}{Mt}
\biggr ) \biggr ).
\end{equation}
and we have assumed $M$ and $m_{0}$ to be comparable.
The effect of changing $\beta_0$ is only visible in the $O(1/Mt)$
term.
In fact, we are being unnecessarily restrictive in wanting to
preserve the idnetical behaviour of (\ref{wc}) and (\ref{wC}).
{}From (\ref{ni}) onwards it follows that the overall scale of $W$ is
immaterial to the vortex density.  All that is required for
identical {\it leading} behaviour in relativistic and non-relativistic
regimes is that ${\cal C}(k, \mu_{0})$ varies
slowly in the vicinity of the peak of the integrand at
$tk_{c}^{2}\simeq M$.  [$A(k)$ is already slowly varying].
This is the case when, allowing for
coefficients $O(1)$,
\begin{equation}
1 < \frac{\mu_{f}}{|\mu_{0}|} = \frac{M^{2}}{m_{0}^{2}}\ll tM <
\ln\biggl (\frac{1}{\lambda}\biggr )
\label{lims}
\end{equation}
where, in the same spirit, we have taken $|\mu_{0}| < \mu_{f}$.  The
upper bound on $tM$ is a reminder that interactions are always
present, and the Gaussian approximation must fail as soon as the
field fluctuations have extended to the true ground-states at the
minima of the potential.  The lower bound is necessary for the
integrand to be peaked strongly so that the saddle-point
approximation is valid.  As long as (\ref{lims}) is basically
correct, any difference between the relativistic and
non-relativistic regimes will be non-leading.

Assuming these limits, we recover what would have been our first naive guess
for a correlation
function based on the growth of the unstable modes,
\begin{equation}
W(r;t)\simeq\int_{|{\bf k}|<M} d \! \! \! / ^3 k
\, e^{i {\bf k} . {\bf r} }
\;e^{2\Omega (k)t},
\label{Wapp}
\end{equation}
We understand the dominance of wavevectors at $k_{c}^{2} = M/t$ in
the integrand as indicating the formation of domains of mean size
\begin{equation}
\xi (t) = O(\sqrt{t/M})
\label{xit}
\end{equation}
 once $Mt > 1$.  Specifically, we take
\footnote{This somewhat arbitrary choice differs from that of \cite{boyanovsky}
and $I$ by a factor of $\sqrt{2}$.}
$\xi (t) = 2 \sqrt{t/M}$.
Once $Mt > 1$ then $\xi (t) >
M^{-1}$, where $M^{-1}$ characterises the cold vortex radius. In the weak
coupling
approximation
individual domains become
large enough to accomodate many vortices before the approximation
breaks down.  There is no difficulty with
causality since domains increase in size as $\dot{\xi} =
\frac{1}{\sqrt{Mt}} < 1$.
On neglecting terms exponentially small in $Mt$, $W(r;t)$ of
(\ref{Wapp}) can be further rewritten as
\begin{eqnarray}
 W(r;t ) & \simeq &
 e^{2Mt}\int_{0}^{\infty} dk\,\, {\rm sinc}(kr)\, k^{2} e^{-tk^{2}/M}
\label{WG}
\\
&=& W(0;t)\,exp\biggl (-\frac{r^{2}}{\xi^{2}(t)}\biggr )
\label{WG2}
\end{eqnarray}
where
\begin{equation}
W(0;t) \approx C \frac {e^{2M t} } {(M t) ^{3/2} },
\end{equation}
for some C. The exponential growth
of $W(0;t)$ in $t$ reflects the way the field amplitudes fall
off the hill centred at $\Phi = 0$.  With the peaking in wavelength $l
= k^{-1}$
understood as indicating the appearence of domains of characteristic
linear dimension $\xi (t)$, the Gaussian in $r$ is a reflection
of the rms variation $\Delta\xi$ in domain size $\xi$.
This variation is large.
If we isolate the Gaussian saddle-point in (\ref{Wapp}) as
\begin{equation}
W(r;t)\simeq  e^{2Mt}\int_{0}^{M} dk\,\, {\rm sinc}(kr)\, k_{c}^{2} e^{-(k -
k_{c})^{2}/2(\Delta k)^{2}}
\label{Wapp2a}
\end{equation}
 then
\begin{equation}
\frac{\Delta\xi}{\xi} = \frac{\Delta k}{k_{c}} = \frac{1}{2}.
\label{rms}
\end{equation}

To calculate the number density of vortices at early times we
insert the expression (\ref{WG}) for $W$ into the equations derived
earlier, to find
\begin{equation}
n(t) = \frac{1}{\pi}
\frac{1}{\xi(t)^2}
\label{nt}
\end{equation}
for an $O(2)$ theory with strings in three dimensions.  We note that the
dependence on time
$t$ of both the density and density correlations
 is only through the correlation length $\xi (t)$.  We have a {\it
scaling} solution in which, as the
domains of coherent field form and expand, the interstring distance
grows accordingly.  Since the only way the defect density can
decrease without the background space-time expanding is by
defect-antidefect annihilation, we deduce that the coalescence of
domains proceeds by the annihilation of small loops of
string.  However, because the density of vortices only depends on $\xi
(t)$ in this early stage, the fraction of string in `infinite'
string remains constant.
Thus, at the same time as small loops disappear,
other loops must rearrange themselves so that the length of
`infinite string decreases accordingly.
Finally, there is roughly
one string zero per coherence area,  a long held belief for whatever
mechanism.

There is one final concern.  A necessary condition for this rolling
down of the field to be valid is that the initial field distribution
at $t \leq 0$ should not overhang the point of inflection at $\langle |\phi
|\rangle =
\sqrt{2M^{2}/ 3\lambda} = O(T_{c})$.  That is,
the initial field fluctuations about $\phi_{a} = 0$ should
be small enough that there is no significant probability that the
field is already in the true vacuum. To
check this for the relativistic regime, we anticipate that, when a domain
structure forms, the
smallest domains that can be identified will be of the size of the
cold vortex radius, the Higgs field Compton wavelength,
$\xi_{0} = O(M^{-1})$.  For temperature $T >
T_{c}$ (but not too close), the rms field fluctuations on this scale
are \cite{ray} $\Delta\phi = O((TM)^{\frac{1}{2}})$.  The condition
that $(\Delta\phi )^{2} < T_{c}^{2}$ is guaranteed when $T_{c}$ is
much larger than $M$, as happens for small coupling.  In fact, since
there are small prefactors, the coupling does not have to be very
small for this to happen. Details are given in $I$.
For the non-relativistic case there are no reasons, a priori, why
the situation should be different.

\section{Vortex Density Correlations}

In addition to the gross vortex density (\ref{nt}) we can calculate
the density-density correlation functions $C_{ij}(r;t)$ of
(\ref{ddc}), identical for both the relativistic  plasma and the
non-relativistic medium.  These were not considered in $I$.

Yet again, as in the case of the density $n(t)$, the $t$-dependence  of
$C_{ij}$ only occurs implicitly through $\xi (t)$.
The simple analytic form of (\ref{WG}) enables us to calculate $A$ and
$B$, up to exponentially small terms in $Mt$, as
\begin{eqnarray}
A(r;t) &=& \frac{2}{\pi^{2}\xi^{4}(t)}
\frac{e^{-2r^{2}/\xi^{2}(t)}}{(1 - e^{-2r^{2}/\xi^{2}(t)})^{2}}
\biggl [(1 - e^{-2r^{2}/\xi^{2}(t)}) - 2\frac{r^{2}}{\xi^{2}(t)}
\biggr ] < 0,
\nonumber
\\
B(r;t) &=& \frac{2}{\pi^{2}\xi^{4}(t)}
\frac{e^{-2r^{2}/\xi^{2}(t)}}{(1 - e^{-2r^{2}/\xi^{2}(t)})} > 0,
\label{corr1}
\end{eqnarray}
whence the diagonal elements $C_{ii}$ (no summation) are all {\it
negative}. Specifically,
\begin{equation}
A(r;t) -   B(r;t) = -\frac{2}{\pi^{2}\xi^{4}(t)}
\biggl (\frac{2r^{2}}{\xi^{2}(t)}\biggr )
\frac{e^{-2r^{2}/\xi^{2}(t)}}{(1 - e^{-2r^{2}/\xi^{2}(t)})^{2}}.
\label{corr2}
\end{equation}
That is, we have {\it anticorrelation} of densities for parallel
directions and positive $B$ for orthogonal ones.  This is just the
situation discussed in Section 2.

It  is useful to expand (\ref{corr1}) and (\ref{corr2}) for $r <
\xi$.  On normalising by a factor of $n^2$, $a$ and
$b$ of (\ref{abc}) are given by
\begin{eqnarray}
a(r;t) &=& -1 + O\biggl (\frac{r^{2}}{\xi^{2}}\biggr ) < 0,
\nonumber
\\
b(r;t) &=& \biggl (\frac{\xi^{2}(t)}{r^{2}}\biggr )
+ O(1) > 0.
\label{corr3}
\end{eqnarray}
For $r > \xi$  there is exponential falloff but, from (\ref{corr3})
we see that, in units of $(\pi\xi^{2})^{-2}$, the
anticorrelation is large.
The discussion of Section 2 is directly relevant.
Since strings with a
long persistence length would imply {\it positive} parallel correlations,
and weak orthogonal correlations, we can
interpret these anticorrelations as a reflection of an increased
string bendiness.
Although it is difficult to be precise, this suggests a significant
amount of string in small loops.
This is an important issue since,
as we noted, early universe
cosmology requires infinite string if string is to be the source
of large-scale structure.  In practice, some is enough.

There  is another, indirect, way in which we can see the tendency for
more string to
be in small loops than we might have thought.
In numerical simulations of string networks based in throwing down
field phases at random,
the rule of thumb for a {\it regular} domain structure in field
phase is that a
significant fraction of string, if not most, is
infinite string \cite{tanmay,mark}.

However, as we saw earlier in (\ref{rms}), we do not have a regular
domain structure in our model but have domains with a large variance
in their size $\Delta\xi /\xi = \frac{1}{2}$.
Unfortunately, the domain structure that we have here is not yet appropriate
for a direct comparison since, as well as the domain size, the field magnitude
has a variance about $\langle |\phi|\rangle = O(M\,e^{Mt})$.  The
work of Guth and Pi \cite{guth} shows that this can be parametrised
by an effective dispersion in $t$ in $\langle |\phi|\rangle$ of
$\Delta t = O(M^{-1})$.  Nonetheless, consider an
idealised case in which domain growth stops instantaneously because
of back-reaction at some time $t_{f}$.  The distribution of strings will then
be as
above for $\xi = \xi (t_{f})$, while the field adjusts to the vacuum manifold,
without
changing phase, in each domain.  We would expect some
string-antistring annihilation to continue while this adjustment
occurs, so that the $n(t)$ calculated previously is an overestimate
of the string density at the end of the transition.  However, the
domains will still be of varying size with variance plausibly given
approximately by (\ref{rms}).  Although it has not been introduced
along the lines above, the inclusion of domain variance in
numerical simulation of string networks shows \cite{andy} that, the greater the
variance, the more string is in small loops.
Beyond observing that there
seems to be some infinite string, we will say no more.

All our results are for quenches that go from significantly above the
transition to significantly below it.  These give the {\it smallest} value
of $\Delta\xi /\xi$ possible and, plausibly, the best chance of
producing infinite string.  Suppose the {\it initial} state
characterised by $S_{0}$ of (\ref{S0})
is very close to the transition.  For relativistic theories this
means starting from a temperature only just above the critical
temperature.  For a non-relativistic theory it corresponds to
beginning from a density very close to the critical density.  In
either case it corresponds to taking $m_{0}^{2}$ suffficiently small
that the inequalities (\ref{lims}) cannot be satisfied.  Some
caution is required, since fluctuations are large near the
transition and the semiclassical action $S_{0}$ of(\ref{S0})
is unreliable.  Despite that, let us take it seriously and assume
that it is possible to neglect $m_{0}^{2}$ in comparision to
$k_{c}^{2}$ within the time interval $1 < Mt < \ln(1/\lambda )$.

The overhang calculation of $I$ shows that there is some leeway. Then, with
\begin{equation}
\frac{T_{0}}{{\bf k}^{2} + m_{0}^{2}}\simeq \frac{T_{0}}{{\bf
k}^{2}}
\end{equation}
 in (\ref{wC}) and (\ref{wc}) no longer slowly varying,
instead of (\ref{Wapp}), we have
\begin{eqnarray}
W(r;t) &\simeq & \int_{|{\bf k}|<M} \frac{d \! \! \! / ^3 k}{{\bf k}^{2}}
\, e^{i {\bf k} . {\bf x} }
\;e^{2\Omega (k)t}
\label{Wapp4}
\\
&\simeq & \int dk\,\, {\rm sinc}(kr)\,\, e^{2\Omega (k)t}
\label{Wapp5}
\end{eqnarray}
up to irrelevant factors.

There is now no peaking of the integrand in (\ref{Wapp5}) at any
preferential wave number, and hence no preferred length scale that
can be identified as a typical domain size.    This lack of
domain structure,
and the inapplicability of the Kibble mechanism, is because of the enhanced
long-wavelength fluctuations that come from being too close to the transition
initially.  We stress that this does not mean that there are no
vortices, or even a small vortex density.
$W(r;t)$ of (\ref{Wapp5}) can be integrated as
\begin{equation}
W(r;t) = W(0;t)\, _{1}F_{1}\biggl (\frac{1}{2},\frac{3}{2},
-\frac{r^{2}}{\xi^{2}(t)}\biggr )
\end{equation}
for $\xi (t)$ as before.
A little algebra shows that the vortex density $n(t)$ is
\begin{equation}
n(t) = \frac{1}{3\pi}\frac{1}{\xi^{2}(t)},
\end{equation}
one-third of its previous value in (\ref{nt}).
For distances $r > \xi$, $W(r;t)$ now has power behaviour
\begin{equation}
W(r;t) = O\biggl (\frac{\xi (t)}{r}\biggr )
\label{WG6}
\end{equation}
instead of the Gaussian behaviour of (\ref{WG2}).
As before, there is anticorrelation with $A < 0, B > 0$ but, at large
distances,
 the density-density correlation
functions are very different, only
showing power-behaviour in their fall-off.
However, for $r\simeq\xi$ their behaviour is very like that of (\ref{corr3}).
Nonetheless, if we
interpret (\ref{Wapp5}) as the limiting case of domain structure with
maximum variance in size (compatible with causality),
the same numerical data quoted earlier
\cite{andy}, with due caveats, suggest even more string in small loops.

Finally, with superfluid films in mind, we can repeat the analysis
in {\it two} space dimensions.
Instead of (\ref{WG2}) $W(r;t)$ is given, when $Mt\ge 1$, by
\begin{equation}
 W(r;t )  \simeq
 e^{2Mt}\int dk\,\, {\rm sinc}(kr)\, k e^{-tk^{2}/M}
\label{WG3}
\end{equation}
The integrand is strongly peaked at $k^{2}_{c} = M/2t$ which, from
our previous analysis, we interpret as the existence of domains,
 trapping $O(2)$ monopoles (vortex cross-sections) on
their  boundaries.
The domains are of varying size.
If we isolate the Gaussian saddle-point in (\ref{WG6}) as
\begin{equation}
W(r;t)\simeq  e^{2Mt}\int_{0}^{M} dk\,\, {\rm sinc}(kr)\, k_{c} e^{-(k -
k_{c})^{2}/2(\Delta k)^{2}}
\label{Wapp2b}
\end{equation}
 then the rms variation $\Delta\xi$ in domain size $\xi$ is
larger than in three dimensions, as
\begin{equation}
\frac{\Delta\xi}{\xi} = \frac{\Delta k}{k_{c}} = \frac{1}{\sqrt{2}}.
\label{rms2}
\end{equation}
This is independent of the definition of $\xi$, but if we take
$\xi^{2}(t) = 4/k^{2}_{c}$ as before, then
\begin{eqnarray}
W(r;t) &=& W(0;t)\, _{1}F_{1}\biggl ( 1, \frac{3}{2},
-\frac{2r^{2}}{\xi^{2}(t)}\biggr )
\label{WG4}
\\
& = & O\biggl (\frac{\xi^{2}(t)}{r^{2}}\biggr )
\end{eqnarray}
for $r > \xi$.
This is intermediate between (\ref{WG6}) and the Gaussian behaviour of
(\ref{WG2}).  Yet again there is anticorrelation in the
density-density correlation function $\langle\rho ({\bf r})\rho
({\bf 0})\rangle = A(r;t)$, where $A(r;t)$ is given in (\ref{defA}).

\section{Conclusions}

In this paper we have shown how global O(2) vortices appear, at a
quench from the ordered to disordered state, as a consequence of the
growth of unstable Gaussian long wavelength fluctuations.  Most
importantly, in the light of discussions about the extent to which
vortex production in low-temperature many-body systems simulates
vortex production in the early universe, our model supports the
analogy.  We have shown how, with our simple assumptions, vortex
production is {\it identical} in both a relativistic
high-temperature quench, in which the initial state is characterised
by $T\gg m$, and in a non-relativistic density quench in which the
initial state is described by $T\ll m$ (to freeze out antiparticles)
with a chemical potential $\mu_{{\rm nr}}\simeq T$.  All  that is
required is an appropriate translation of the parameters from the
one case to the other.

In our simple model of Gaussian fluctuations the resulting string
configurations scale as a function of the correlation length $\xi
(t) = O(t^{\frac{1}{2}})$, at about one vortex/correlation area.
This is compatible with the Kibble mechanism for vortex production
on domain boundaries upon phase separation.
For a weak coupling theory the domain cross-sections are
significantly larger than a vortex cross-section at the largest
times for which the approximations are valid.  However,  there is a large
variance in their size, with $\Delta\xi /\xi = \frac{1}{2}$.
Because of this there is more string in
small loops than we might have anticipated.
Nonetheless, we expect some infinite string.

We stress that our model can, at best, describe weak coupling
systems for the short times while the domains are growing
 before the defects freeze out.  This is
unsatisfactory for most early universe applications and for
low-temperature many-body systems.  However, we know in principle
\cite{boyanovsky} how to include back-reaction (still within the
context of a Gaussian approximation) to slow down domain growth as
the field fluctuations spread to the ground-state manifold.  The
identity of the weak-coupling results of the two regimes should
survive to this case also, although it will probably lead to different
conclusions from those above.  Further, there is no difficulty in
principle of embedding these results in a FRW metric, along the
lines of \cite{hector}.

\section*{Acknowledgements}

The authors are grateful to T.W.B.\
Kibble for useful conversations on defect formation.
Two of us (R.J.R.\ and A.J.G.) would like to thank Prof. M. Krusius and
Prof. G. Volovik for their hospitality
at the Low Temperature Laboratory, Helsinki University of
Technology, where some of this work was performed.
T.S.E.\ wishes to thank the Royal Society for its support and
A.J.G.\ would like to thank PPARC.

\renewcommand{\theequation}{A\arabic{equation}}
\setcounter{equation}{0}
\section*{Appendix}

We choose to work with `physical' fields, that is fields which
evolve according to the original Hamiltonian, $H_{0}$, which has not had
the chemical potential absorbed into it.  The chemical potential
then appears in the boundary conditions and so is placed on the
same footing as the temperature.  For this reason we work with the
eigenstates of $Q$, $\phi (x)$ and $\phi^\dagger  (x)$, rather than
the Cartesian components $\phi_{1}(x)$ and $\phi_{2}(x)$.  The
conversion to the Cartesian thermal Wightman functions $W_{ab}(r,
t)$ of the text is trivial.

For the complex scalar field we are interested in, the Hamiltonian has the
usual form as given in (\ref{Hc2}) in which, for the reasons given
in text, $\lambda = 0$.  That is, we are interested in 'free' thermal
Wightman functions, albeit with changing mass.  The Wightman
functions in question are defined to be
\begin{eqnarray}
G^> (x,x') &=& {\rm Tr} \{ e^{-\beta(H_0-\mu Q)} \phi(x) \phi^\dagger  (x') \}
\\
G^< (x,x') &=& {\rm Tr} \{ e^{-\beta(H_0-\mu Q)} \phi^\dagger  (x') \phi(x)
\} .
\end{eqnarray}
Other types of propagator are defined in a similar way,
retarded, advanced, time-ordered etc.\ but will not be needed here
and in any case they can be built from $G^{>(<)}$ relatively easily.

The free propagators satisfy differential equations
which look like the usual Klein-Gordon
equation, the only difference in our case is that the mass term is
time dependent - a simple step function for the quench at time $t =
0$ that we have considered in the text.  Otherwise our dicsussion is
general.  Adopting it to our specific model is straightforward.

We start by solving the homogeneous equation
\begin{eqnarray}
(\partial_t^2+\omega^{2}(t))U^{\pm}(t) &=& 0
\end{eqnarray}
where
\begin{equation}
\omega(t)=\theta(-t) \omega_1 + \theta(t) \omega_2 .
\end{equation}
As it stands both $\omega_1$ and $\omega_2$ are assumed to be real.
If we have an unstable mode at late times then we just replace
$\omega_2= \pm i \Omega$.
The boundary conditions are
\begin{eqnarray}
\lim_{t \rightarrow - \infty} U^{\pm}(t) &=& \exp \{  \pm i \omega_1 t
\}
\\
U^\pm(t=0^+) = U^\pm(t=0^-) &&
\partial_t U^\pm(t=0^+) = \partial_t U^\pm(t=0^-)  .
\end{eqnarray}
This gives
\begin{eqnarray}
U^\pm(t) &=& \theta(-t) e^{  \pm i \omega_1 t }
+ \theta(t) [ A_\pm e^{  \pm i \omega_1 t }
+ B_\pm e^{  \mp i \omega_1 t }  ] \\
&& A_\pm = \frac{1}{2} \left( 1 + \frac{\omega_1}{\omega_2} \right)
\; , \; \; \; \;
B_\pm = \frac{1}{2} \left( 1 - \frac{\omega_1}{\omega_2} \right) .
\end{eqnarray}

Now we can construct the thermal Wightman functions.  As they satisfy
\begin{eqnarray}
(\partial_t^2    + \omega^2(t)) G^{> (<)} (t,t') &=& 0 \\
(\partial_{t'}^2 + \omega^2(t')) G^{> (<)} (t,t') &=& 0
\end{eqnarray}
we try
\begin{equation}
G^{> (<)}(t,t') = a^{> (<)} U^-(t')U^+(t) + d^{> (<)} U^+(t') U^-(t)
\end{equation}
where terms such as $U^-(t')U^-(t)$ and $U^+(t')U^+(t)$ are
eliminated by demanding time translation at early times.

We start by noting that it is common, when considering the thermal
Wightman functions of real fields in thermal equilibrium, to use their
definitions to deduce a
trivial identity relating the two thermal Wightman functions at
equal times.  In the case of the complex fields we see that we have
\begin{eqnarray}
[G^{> (<)} (t,t')]^\ast = G^{> (<)} (t'{}^\ast,t^\ast) &&
G^{> (<)} (t,t) = G^{> (<)} (t^\ast,t^\ast)  .
\end{eqnarray}
We have been very careful to ensure that the definition of
$\phi^\dagger(t)$ is such that it evolves like $\phi(t)$, namely
$\phi^\dagger(t)=\exp\{-iH_{0}t \} \phi^\dagger(0) \exp \{ i H_{0}t
\}$ (where we have dropped spatial labels).
Problems occur with complex time shifts, such as are encountered
in thermal field theory, as roughly speaking we require
$(\phi(t))^\ast = \phi^\dagger(t^\ast)$.  The $U^+$ and $U^-$ functions below
are related in a similar way.

The ETCR (equal time commutation relations) can be used to provide
boundary conditions,
\begin{equation}
[\phi(t), \Pi(t)] = [\phi^\dagger (t),\Pi^\dagger (t)] = i .
\end{equation}
We have $\Pi(t)=\partial_t \phi^\dagger (t)$ and $\Pi^\dagger (t)=
\partial_t \phi(t)$ since, for the physical fields,
there is no chemical potential in the expressions for $\Pi$ and $\Pi^\dagger$.
Looking at these equations for any one negative time gives
\begin{eqnarray}
a^< - d^< - a^> + d^> &=& \frac{1}{\omega_1}
\end{eqnarray}
and we find the same relation if the ETCR is imposed at any
positive time.  This confirms that the ETCR are indeed maintained at
all times if they are true at any one time.

There is a second ETCR which we can use;
\begin{equation}
[\phi(t),\phi^\dagger (t)]=0
\end{equation}
This requires
\begin{equation} G^>(t,t)=G^<(t,t) \label{grel2} \end{equation} which gives
\begin{equation}
a^> + d^> = a^< +d^< .
\end{equation}
Once this is satisfied it ensures that this ETCR also holds at all times.

The temperature and chemical potential appear only in the last
boundary condition.  This is the well known KMS condition \cite{LvW}.
Here we enforce this at any negative time as then the system is in
equilibrium.  Thus
\begin{equation}
G^>(t,t') = e^{-\beta \mu } G^<(t+i\beta,t')
\end{equation}
This holds for all $t,t'<0$, but not for later times,
so we find
\begin{eqnarray}
a^> &=& e^{-\beta(\omega_1 + \mu)} a^<
\\
d^> &=& e^{\beta(\omega_1 - \mu)} d^<
\end{eqnarray}
Putting all this together gives
\begin{eqnarray}
a^< &=& \frac{1}{2 \omega_1} \frac{ 1}{1-e^{-\beta(\omega_1+\mu)} }\\
a^> &=& \frac{1}{2 \omega_1} \frac{-1}{1-e^{ \beta(\omega_1+\mu)} }\\
d^< &=& \frac{1}{2 \omega_1} \frac{-1}{1-e^{ \beta(\omega_1-\mu)} }\\
d^> &=& \frac{1}{2 \omega_1} \frac{ 1}{1-e^{-\beta(\omega_1-\mu)} }
\end{eqnarray}

For our purposes we just need the equal time propagator which can be
obtained from either of the thermal Wightman functions because of the
second ETCR (\ref{grel2}).  From the above we find that
\begin{equation}
G^>(t,t)= G^<(t,t)
=\frac{1}{2\omega_1}
\frac{e^{-\beta \omega_1} - e^{\beta \omega_1}}{
e^{\beta \mu} - e^{-\beta \omega_1} - e^{\beta \omega_1} + e^{-\beta \mu}}.
U^-(t)U^+(t)
\end{equation}
from which our results of section 5 follow.

We might ask whether we need worry about {\it off-diagonal} terms in
the Cartesian propagator or, equivalently,
$\langle \phi(t) \phi(t') \rangle$ and
$\langle \phi^\dagger(t) \phi^\dagger(t') \rangle$.
In some approaches using different definitions of the
fields, e.g.\ \cite{dodelson}, these are not zero.
Trying the same ansatz as before, we see that
time-translation invariance and the KMS condition, applied at
negative times, ensure that these
are zero for all times in our case.
This is a big advantage of our definitions.


\begin{thebibliography}{99}

\bibitem{kibble1} T.W.B. Kibble, J. Phys. {\bf A9}, 1387 (1976).

\bibitem{shellard} E.P.S Shellard and A. Vilenkin, {\it Cosmic Strings
and other Topological Defects}, Cambridge University Press (1994)


\bibitem{lancaster} P.C. Hendry, N.S. Lawson, R.A.M. Lee, P.V.E. McClintock
and C.D.H. Williams, Nature 368, 315 (1994).

\bibitem{helsinki} V.M.H. Ruutu, M. Krusius, U. Parts,
G.E. Volovik,  {\it Neutron Mediated Vortex Nucleation in
Rotating Superfluid $^{3}He-B$}, (Low Temperature Laboratory, Helsinki
University of
Technology, 02150 Espoo, Finland), in preparation.

\bibitem{lancaster2} P.C. Hendry, N.S. Lawson, R.A.M. Lee, P.V.E. McClintock
and C.D.H. Williams, Newton Proceedings.


\bibitem{zurek1} W.H. Zurek, Nature {\bf 317}, 505 (1985),
Acta Physica Polonica, {\bf B24}, 1301 (1993).

\bibitem{zurek2} W.H. Zurek, to be published in the Proceedings of the NATO
Advanced Study Institute on {\it Topological Defects}, Newton Institute,
Cambridge
1994, (Plenum Press, to appear).

\bibitem{kleinert} H. Kleinert, {\it Gauge Fields in Condensed
Matter, Vol.I}, World Scientific (1988).

\bibitem{tanmay} T. Vachaspati and A. Vilenkin, Phys. Rev. {\bf D30},
2036 (1984).

\bibitem{alray} A.J. Gill and R.J. Rivers,
{\it Phys. Rev.}{\bf D51} (1995) 6949.

\bibitem{timray} R.J. Rivers and T.S. Evans, {\it The Production of
Strings and Vortices at Phase Transitions},
to be published in the Proceedings of the NATO
Advanced Study Institute on {\it Topological Defects}, Newton Institute,
Cambridge
1994, (Plenum Press, to appear).


\bibitem{halperin} B.I. Halperin, {\it Statistical Mechanics of
Topological Defects}, published in {\it Physics of Defects},
proceedings of Les Houches, Session XXXV 1980 NATO
ASI, editors Balian, Kl\'{e}man and Poirier (North-Holland Press)
816 (1981).

\bibitem{schwinger} J. Schwinger, J. Math. Phys. {\bf 2}, 407 (1961).

\bibitem{mahantapa} K.T. Mahanthappa and P.M. Bakshi, J.M. Phys. {\bf
4}, 1; ibid, 12 (1963).

\bibitem{keldysh} L.V. Keldysh, Sov. Phys. JETP {\bf 20}, 1018 (1965).

\bibitem{Go4}  K-C. Chou, Z-B. Su, B-L. Hao and L. Yu, Phys.
Rep. {\bf 118} (1985) 1.

\bibitem{altimray2} A.J. Gill, T.S. Evans and R.J. Rivers, {\it
The Chemical Potential and the Non-Relativistic Limit},
Imperial College Preprint (in preparation).

\bibitem{ray2} R.J. Rivers {\it Vortex Production at Phase
Transitions in Non-Relativistic and Relativistic Media}, to be
published in the proceedings of the {\it 3me. Colloque Cosmologie,
Paris, June 1995} (World Scientific Press) 1996.

\bibitem{ray3} R.J. Rivers{\it The Similarity of Vortex Production at Phase
Transitions in Non-Relativistic and Relativistic Media}, to be
published in the proceedings of the {\it 4th Thermal Fields Workshop,
Dalian, China, August 1995} (World Scientific Press) 1996.

\bibitem{guth} A. Guth and S-Y. Pi, Phys. Rev. {\bf D32}, 1899 (1985).

\bibitem{weinberg} E.J. Weinberg and A. Wu, Phys. Rev. {\bf D36},
2474 (1987).

\bibitem{boyanovsky} D. Boyanovsky, Da-Shin Lee and Anupam Singh,
Phys. Rev. {\bf D48}, 800 (1993).

\bibitem{kapusta} J.I. Kapusta, {\it Finite Temperature Field
Theory}, Cambridge University Press, (1989).
\\
J.I. Kapusta, Phys. Rev. {\bf D24}, 426 (1981)

\bibitem{art} H.E. Haber and H.A. Weldon, Phys. Rev. {\bf D25}, 502
(1982).

\bibitem{dodelson} J. Bernstein and S. Dodelson, Phys. Rev. Lett.
{\bf 66}, 683 (1991).
\\
K.M. Benson, J. Bernstein and S. Dodelson, Phys. Rev.
{\bf D44}, 2480 (1991).

\bibitem{wiegel} F.W. Wiegel, {\it Introduction to Path Integral
Methods in Physics and Polymer Science}, (World Scientific Press), (1986).

\bibitem{ray} M. Hindmarsh and R.J. Rivers, Nucl. Phys. {\bf 417},
506 (1994).

\bibitem{mark} M. Hindmarsh and K. Strobl, Nucl. Phys. {\bf B437},
471 (1995).

\bibitem{andy} A. Yates, Imperial College preprint (in preparation).

\bibitem{hector} D. Boyanovsky, H.J. de Vega, R. Holman, Phys. Rev.
{\bf D49}, 2769 (1994)

\bibitem{LvW} N.P. Landsman and Ch.G. van Weert,
Phys. Rep. {\bf 145} 141 (1987).

\end{thebibliography}
\end{document}